\documentclass{aa}  

\usepackage{graphicx}
\usepackage{txfonts}
\usepackage{amsmath}
\usepackage{amssymb}
\usepackage{natbib}
\usepackage{dblfloatfix}
\usepackage{placeins}

\graphicspath{{./figures/}}
\bibpunct{(}{)}{;}{a}{}{,}
\newlength{\minuslength}
\begin{document}

   \title{A tale of two isotopes}

   \subtitle{Spatial variation in HCN fractionation toward young cores}

   \author{S. S. Jensen\inst{1}\thanks{\email{sigurdsj@mpe.mpg.de}}
                \and S. Spezzano \inst{1}
                \and O. Sipil{\" a}\inst{1}
                	\and P. Caselli\inst{1}
                	\and L. Colzi\inst{2}
                \and E. Redaelli \inst{3}
        }

   \institute{Max-Planck-Institut f{\"u}r extraterrestrische Physik, Giessenbachstrasse 1, D-85748 Garching, Germany
	\and Centro de Astrobiolog{\' i}a (CAB), CSIC-INTA, Carretera de Ajalvir km 4, Torrej{\'o}n de Ardoz, 28850 Madrid, Spain
	\and European Southern Observatory, Karl-Schwarzschild-Strasse 2, D-85748 Garching, Germany}

   \date{Draft date: \today}

  \abstract
  {Isotopic fractionation can serve as a powerful tracer of the chemical evolution during star and planet formation. To accurately interpret observations, it is crucial to identify the dominant pathways of nitrogen and carbon fractionation at different evolutionary stages.}
  {We aim to study nitrogen and carbon fractionation in a sample of young cores at the onset of star formation.}
  {We map H$^{13}$CN and HC$^{15}$N around one starless and three pre-stellar cores. We compute the $N$(H$^{13}$CN)/$N$(HC$^{15}$N) column density ratio across the cores and compare the distribution with $N$(H$_2$) maps from \emph{Herschel}/SPIRE. In addition, we calculate $^{14}$N/$^{15}$N maps using the double isotope method for comparison with earlier studies. The results are compared with astrochemical modeling of carbon and nitrogen fractionation for a one-dimensional pre-stellar core model.}
  {The computed $N$(H$^{13}$CN)/$N$(HC$^{15}$N) ratio exhibit clear spatial variation across the maps. This variation is correlated with $N$(H$_2$) in three out of four cores.}
  {Our analysis reveals a correlation between the H$^{13}$CN/HC$^{15}$N ratios and the $N$(H$_2$) maps. According to the astrochemical model, the correlation is mainly due to variations in the $^{12}$C/$^{13}$C ratio. Consequently, the results caution against applying the double-isotope method to derive $^{14}$N/$^{15}$N ratios without independently assessing possible spatial variations in the $^{12}$C/$^{13}$C ratio. Furthermore, the leading cause of the isotopic variation in the model is not isotope-selective photodissociation, but rather more efficient fractionation through exchange reactions at lower temperatures in the denser regions of the cores.}


   \maketitle
   \nolinenumbers
%
\section{Introduction}
Isotopic fractionation is one of the principal tools in astrochemical studies of chemical evolution during star and planet formation \citep[e.g.,][]{2012A&ARv..20...56C, 2023ASPC..534.1075N}.
 Among the prominent fractionation systems, nitrogen fractionation ($^{14}$N/$^{15}$N) and carbon fractionation ($^{12}$C/$^{13}$C) are of particular interest for several reasons. First, both nitrogen and carbon are amongst the most abundant elements, which means that molecules containing nitrogen and carbon are abundantly available at each stage of star and planet formation, making them strong tracers of chemical evolution \citep[e.g.,][]{2011ARA&A..49..471M, 2017A&A...603L...6H, 2019A&A...632L..12H, 2023A&A...674L...8R}. Second, both elements are key to the Earth's biosphere: nitrogen is the main constituent of the atmosphere and crucial for synthesizing prebiotic molecules, including RNA \citep{2020ApJ...899L..28R, 2020ChRev.120.4616S}, while carbon is the backbone of organic chemistry. Third, the $^{14}$N/$^{15}$N ratio is also an important tracer of nucleosynthesis on Galactic scales \citep[e.g.,][]{2022A&A...667A.151C}. 
 
 Despite its numerous applications, the fundamental pathways to nitrogen and carbon fractionation remain uncertain. Like other isotopic fractionation systems, nitrogen fractionation can occur through exchange reactions in the cold phases of the interstellar medium (ISM) \citep[e.g.,][]{1981ApJ...247L.123A, 2015A&A...576A..99R}. However, additional processes -- such as nucleosynthetic variations and isotope-selective photodissociation -- may also contribute substantially in shaping the complex pattern of nitrogen fractionation observed in star-forming regions and in the Solar System \citep{2012ceg..book.....M, 2014A&A...562A..61H, 2019MNRAS.490.2838R}. Similarly, carbon fractionation can occur through imbalances in chemical exchange reactions and may also be influenced by isotope-selective photodissociation \citep{1982ApJ...255..143B, 2009A&A...503..323V}.
 
 Isotope-selective photodissociation of nitrogen occurs because N$_2$ undergoes effective self-shielding \citep{2014A&A...562A..61H, 2018ApJ...857..105F}. At moderate column densities, the self-shielding of N$_2$ reduces the impact of photochemistry, whereas the rarer isotopolog $^{15}$N$^{14}$N attains effective self-shielding only at higher column densities due to its lower abundance.
 Consequently, photodissociation of $^{15}$NN in denser regions of the ISM can decrease the atomic $^{14}$N/$^{15}$N ratio in the gas-phase and in molecules formed within these regions. Similarly, $^{12}$CO self-shields at lower column densities with respect to $^{13}$CO.
 
It is an open question whether isotope-selective photodissociation plays a major role in the fractionation of nitrogen and carbon or whether other pathways dominate the fractionation process. \citet{2022A&A...664L...2S} recently reported observational evidence of isotope-selective photodissociation of nitrogen in the pre-stellar core L1544. That study was the first to show this effect in a dynamically evolved core. However, astrochemical models have suggested that the observed trends could also result from variations in the $^{12}$C/$^{13}$C ratio across the core \citep{2023A&A...678A.120S}. Therefore, further study of the spatial variation in the nitrogen and carbon fractionation in young cores is needed.

At present, the main molecular carriers of nitrogen during the early stages of star formation remain unknown. Fractionation studies suggest that several reservoirs with distinct fractionation levels may contribute to the present-day distribution, both in the Solar System and in protoplanetary disks \citep{2017A&A...603L...6H, 2019A&A...632L..12H}. To identify the principal nitrogen carriers during star and planet formation, we need a better understanding of the processes that drive nitrogen fractionation: are different fractionation levels the result of chemical reactions or of environmental influences, such as isotope-selective photodissociation?

In this work, we present observations of H$^{13}$CN and HC$^{15}$N toward four young cores. We look for evidence of isotope-selective photodissociation by deriving the HCN fractionation ratios at multiple regions of the core, with varying column densities and inferred visual extinctions. We compare the results with an astrochemical model including both $^{14}$N/$^{15}$N and $^{12}$C/$^{13}$C fractionation networks and the effects of isotope-selective photodissociation. The astrochemical model is run on a pre-stellar core structure to assess how the fractionation ratios vary with radius and how this may impact the observations reported here.

\section{Observations} \label{sec:2}

\begin{table*}
\centering\caption{Summary of the sources observed.}             
\label{table:observations}
\centering          
\begin{tabular}{l c c c c l }  
\hline \hline       
            \noalign{\smallskip}
Source & R.A. (J2000) & DEC (J2000) & $\varv_\mathrm{LSR} $(km/s) & Isolated/Clustered & Starless/Pre-stellar  \\  
            \noalign{\smallskip}
\hline                   
            \noalign{\smallskip} 
            L1521E  & 04:29:15.7 & 26:14:05.0 &  6.7 & Isolated & Starless \\
            HMM--1 & 16:27:58.3 & -24:33:42.0 &  4.3 & Clustered & Pre-stellar  \\
            L429  & 18:17:06.4 & -08:14:00.0  & 6.7 & Isolated & Pre-stellar \\
            L694--2  & 19:41:04.5 & 10:57:02.0  & 9.6 & Clustered & Pre-stellar  \\

            \noalign{\smallskip} 
\hline                                    

\end{tabular}
\tablefoot{Isolated refers to the characterization of the local cloud environment. The evolutionary classification follows \citet{2005ApJ...619..379C}.}
\end{table*}
%
%
%
%

The targets presented here include the starless core L1521E, and three pre-stellar cores: L429, L694-2, and HMM-1. The classification of the cores is based on \citet{2005ApJ...619..379C}. The four sources are located in diverse environments: L1521E is located in a quiescent part of the Taurus molecular cloud, without nearby sources of incident radiation. L694--2 and L429 are both considered isolated cores associated with the Aquila Rift, and HMM-1 is located in a dynamic star-forming environment in the Ophiuchus molecular cloud \citep{2019A&A...630A.136N, 2020A&A...643A..60S, 2023A&A...670A.141T, 2025A&A...702A.210R, 2025A&A...698A.278T}. In this context, `isolated' cores are sources for which no nearby objects have been identified, which could affect the core either through strong irradiation or dynamical perturbations. To the west of HMM-1, an OB association including $\rho$ Oph may increase the irradiation. An overview of the sources is presented in Table \ref{table:observations}.

All sources have been observed with the \emph{Herschel Space Observatory}/SPIRE  instrument. From the SPIRE continuum maps at 250~$\mu$m, 350~$\mu$m, and 500~$\mu$m, H$_2$ column density maps have been derived toward each of the sources by fitting the spectral energy distribution of the cores \citep{2020A&A...643A..60S}.

The data presented in this study were obtained with the Eight MIxer Receiver (EMIR) E090 instrument on IRAM 30m telescope at Pico Veleta. We used the Fourier Transform Spectrometer (FTS) with
a resolution of 50 kHz, achieving a spectral resolution of $\sim0.17$~km~s$^{-1}$ at 86~GHz. Each source was mapped over a region of 2.5$^{'}$x2.5$^{'}$ using On-the-fly (OTF) mapping with position switching. The receiver was tuned to observe the H$^{13}$CN (1--0) and HC$^{15}$N (1--0) transitions in a single spectral setup.

\begin{table}
\centering\caption{Excitation temperatures derived from fits of the H$^{13}$CN 1--0 hyperfine lines toward the brightest position for each source.}             
\label{table:Tex}
\centering          
\begin{tabular}{l l}  
\hline \hline
            \noalign{\smallskip} 
Source & $T_\text{ex}$ (K) \\ 
            \noalign{\smallskip} 
\hline
L429 & 3.3$\pm0.1$ \\
L694--2 & 4.1$\pm0.1$ \\
L1521E & 3.2$\pm$0.1\\
HMM--1 &3.7$\pm0.1$\\
\hline
\end{tabular}
\tablefoot{The hyperfine lines were fitted using {\sc class} HFS fitting tool.}
\end{table}

\begin{table*}
\centering\caption{Molecular transition data for HC$^{15}$N 1--0 and H$^{13}$CN 1--0. Spectral data extracted from CDMS.}             
\label{table:transitions}
\centering          
\begin{tabular}{l c c c c}  
\hline \hline
            \noalign{\smallskip} 
Transition & Freq. (GHz) & $E_{\text{up}}$ (K) &  $g_{\text{up}}$ & $\log$(A$_{ji}$)  (s$^{-1}$) \\ 
            \noalign{\smallskip} 
\hline
HC$^{15}$N J=1--0 & 86.05497 & 4.13 & 3  & -4.65693  \\ 
H$^{13}$CN J=1--0,F=2--1 & 86.33874 & 4.14 & 3 & -4.65258 \\
H$^{13}$CN J=1--0,F=1--1 & 86.34017 & 4.14 & 5 & -4.65262 \\ 
H$^{13}$CN J=1--0,F=0--1 & 86.34225 & 4.14 & 1 & -4.65254 \\ 
\hline
\end{tabular}
\end{table*}

The observations were carried out over three observing sessions in varying weather conditions. An rms noise level of 20-25~mK was achieved for all maps. The observations were reduced using the {\sc class} software as part of the GILDAS software suite \footnote{\url{https://www.iram.fr/IRAMFR/GILDAS}}. Gridding was performed with a pixel size of 5$^{''}$, corresponding to 5 pixels per beam at 3~mm.
Pointing calibrations were carried out every $\sim$1.5 hours, and the focus was checked every $\sim$3 hours. 

\section{Results} \label{sec:3}

\subsection{Observational data}
\begin{figure*}[ht]
\resizebox{\hsize}{!}
        {\includegraphics{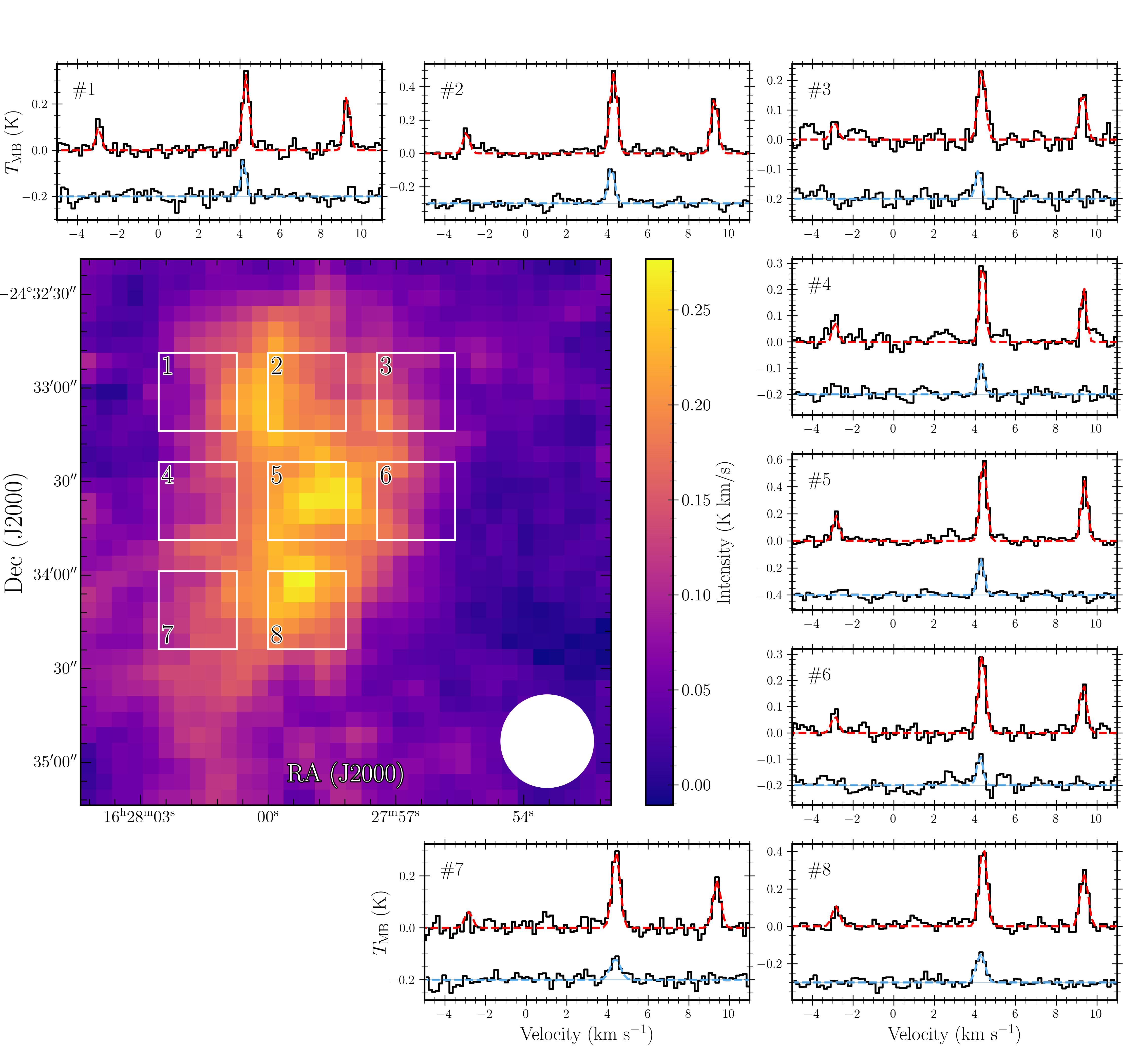}}
  \caption{Moment-zero map toward HMM--1 for the central hyperfine component of H$^{13}$CN (1--0). White rectangles indicate the regions over which the spectra are averaged. The H$^{13}$CN and HC$^{15}$N spectra for each of the 8 positions are plotted in black, the latter offset for clarity. The hyperfine fit for H$^{13}$CN is shown in red, and the Gaussian fit for HC$^{15}$N is shown in blue. The beam size is shown in the lower right corner.}
     \label{fig:HMM-1}
\end{figure*}

Figure \ref{fig:HMM-1} shows the integrated intensity map of H$^{13}$CN (1--0) for HMM--1 and the extracted spectra towards 8 positions in the map. For H$^{13}$CN (1--0), the hyperfine structure of the transition is resolved, and the central, brightest component is used for the integration. Similar figures for the remaining three cores are shown in Appendix \ref{app:figs}. 

\begin{figure*}[ht]
\resizebox{\hsize}{!}
{\includegraphics{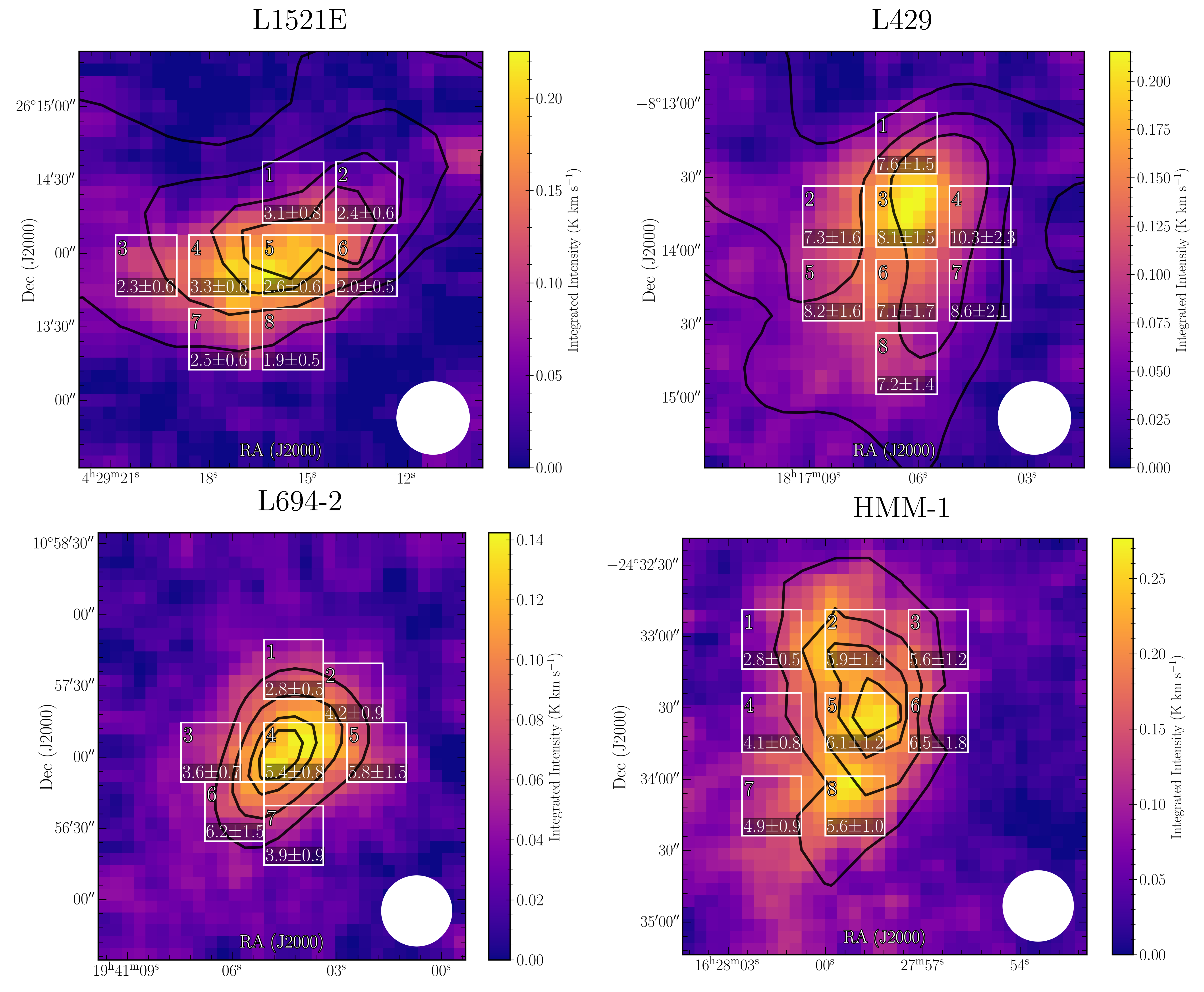}}
  \caption{Derived H$^{13}$CN/HC$^{15}$N column density ratios. Black contours indicate the 10$^{22}$~cm$^{-2}$, $2.5\times$10$^{22}$~cm$^{-2}$, and $5\times$10$^{22}$~cm$^{-2}$ levels in the H$_2$ column density maps derived from \emph{Herschel}/SPIRE observations. The colormap shows the integrated emission of the central hyperfine transition for H$^{13}$CN. For L694--2, positions P2 and P6 are shifted closer to the core center to achieve sufficient S/N.}
     \label{fig:ratios_maps_R2}
\end{figure*}

To derive the column densities of H$^{13}$CN and HC$^{15}$N, we selected between five and nine positions in each map (see Fig. \ref{fig:ratios_maps_R2}). For each position, the spectra were averaged over a 5$\times$5 pixel region, comparable to the beam size, to improve the signal-to-noise ratio (S/N). In doing so, all positions achieved a peak S/N $\geq 5$.
The excitation temperature was obtained by fitting the resolved hyperfine 1--0 transition of H$^{13}$CN. The fit was performed using the {\sc class} HFS fitting method. We utilized the excitation temperature derived for the position with the highest S/N ratio. A common excitation temperature was assumed for both isotopologs, motivated by the assumption of co-spatiality and the close similarities between the transition characteristics for the two isotopologs. In addition, we adopted a single excitation temperature across the maps, similar to \citet{2022A&A...664L...2S}. Moreover, we performed an analysis in which the excitation temperature was derived for each position individually. This approach did not alter any of the trends discussed in this work; the isotopic ratios derived with a uniform excitation temperature differ by $\sim$5--10\% from those obtained using position-dependent excitation temperatures.
The derived excitation temperatures are listed in Table \ref{table:Tex}.

The spectroscopic parameters are listed in Table \ref{table:transitions}, originally provided by \citet{2004ZNatA..59..861F} and retrieved from the CDMS catalog \citep{2001A&A...370L..49M, 2016JMoSp.327...95E}\footnote{\url{https://cdms.astro.uni-koeln.de/}}.

Column densities were derived following:
\begin{equation}\label{eq:1}
	N_\mathrm{tot} = W Q(T_\mathrm{ex}) \frac{8\pi \nu^3}{A_\mathrm{ul} g_\mathrm{u} c^3 } \frac{\exp (\frac{E_\mathrm{u}}{k T_\mathrm{ex}})}{J(T_\mathrm{ex}) - J(T_\mathrm{bg})} \frac{\tau}{1-\exp(-\tau)},
\end{equation}
where $A_\mathrm{ul}$ is the Einstein coefficient for the transitions from upper state $u$ to lower state $l$, $g_\mathrm{u}$ is the degeneracy of the upper state, $Q$ is the partition function, $W$ is the integrated line intensity, $c$ is the speed of light, $\nu$ is the transition frequency, $h$ is the Planck constant, and $k$ is the Boltzmann constant. 
The last factor, $\frac{\tau}{1-\exp(-\tau)}$ corrects for the optical depth \citep{1999ApJ...517..209G}.
The optical depth $\tau$ was derived for the individual components as:
\begin{equation}
	\tau =  \ln \Bigg(\frac{J(T_\mathrm{ex}) - J(T_\mathrm{bg})}{J(T_\mathrm{ex}) - J(T_\mathrm{bg}) - T_\mathrm{mb}}\Bigg )
,
\end{equation}
where $J(T) = \frac{h\nu}{k} (\exp(\frac{h\nu}{kT}) - 1)^{-1}$ is the Rayleigh-Jeans equivalent temperature.

For H$^{13}$CN, we selected the weakest of the hyperfine lines with a peak S/N $> 7.5$ and computed the column density based on this component , instead of using the hyperfine structure fit. This approach was motivated by the high optical depth of the brightest component in the hyperfine ($\tau \gtrsim 1$). Using this approach, the optical depths of the fitted lines for both H$^{13}$CN and HC$^{15}$N are in the range $\tau \leq 0.6$, except for L1521E. An overview of the derived optical depths and column densities toward each position is provided in Appendix \ref{app:tables}.

Figure \ref{fig:ratios_maps_R2} shows maps of the derived H$^{13}$CN/HC$^{15}$N ratio for HCN at each position toward the four sources. The black contours in the figure indicate the $N$(H$_2$) levels derived from \emph{Herschel}/SPIRE, presented in \citet{2020A&A...643A..60S}. 

\begin{figure*}[ht]
\resizebox{\hsize}{!}
        {\includegraphics{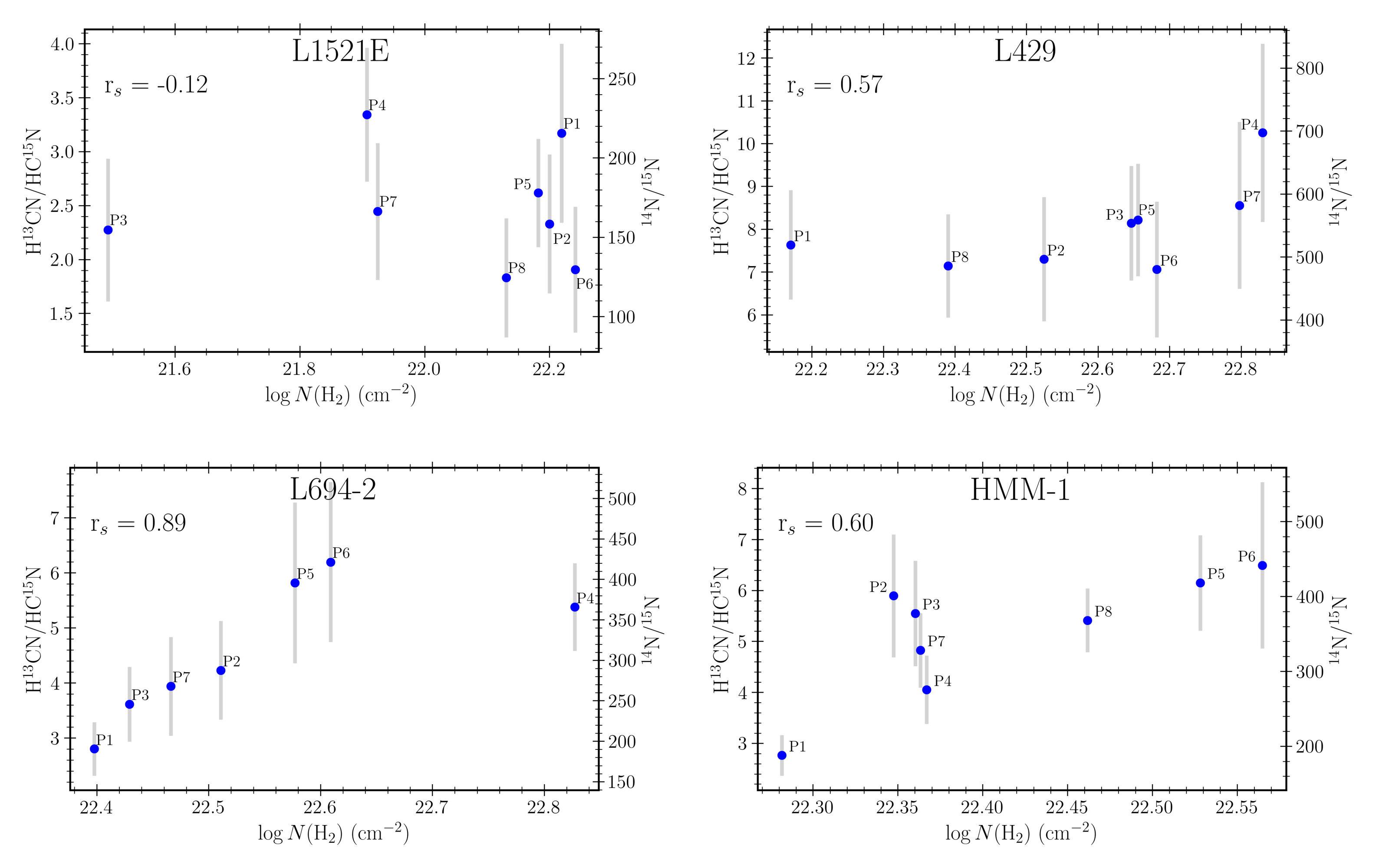}}
  \caption{The H$^{13}$CN/HC$^{15}$N column density ratio (left y-axis) and inferred $^{14}$N/$^{15}$N ratio (right y-axis) as a function of $N$(H$_2$) derived from the \emph{Herschel}/SPIRE maps. The $^{14}$N/$^{15}$N ratios were calculated by naively assuming a constant $^{12}$C/$^{13}$C = 68 across the maps. The computed Spearman correlation coefficients are included in the top-left corner. Points are denoted according to the numbers in Figures \ref{fig:ratios_maps} and \ref{fig:ratios_maps_R2}.}
     \label{fig:scatter}
\end{figure*}

In the maps, the fractionation ratios show considerable spatial variation. Toward L694--2, the nitrogen fractionation ratio is highest at the center of the core (P3) and lowest in positions P1, P2, and P7 at the edge of the core. Positions P5 and P6 also show higher fractionation ratios. The latter two positions coincide with an enhanced H$_2$ column density (see Fig. \ref{fig:scatter}).
 For L1521E, a similar trend is observed, with higher fractionation ratios in the center of the core. The H$_2$ column density drops sharply on the south side of the core (points P7, P8), which would expose this side of the core to greater irradiation from external sources. This coincides with the points exhibiting the lowest fractionation ratios (P6--P8). Conversely, in the central region and on the northern border (P1, P4, P5), fractionation ratios are higher.
 For L429 and HMM-1, patterns are weaker. Nonetheless, the lowest fractionation ratios toward L429 are derived toward the south of the core (P6, P8), which is more exposed to external radiation, as indicated by the $N$(H$_2$) column density maps. 
Similarly, for HMM-1, lower fractionation ratios are again seen on the edge of the core (P1, P3, P4, and P7) compared to the central position (P5).
  
 Overall, the patterns observed in the maps are broadly consistent with the expectations for isotope-selective photodissociation. Nevertheless, the observed patterns may also arise from higher local densities, which influence the fractionation process through enhanced isotopic exchange reactions.
  
To quantify the trends observed in the maps presented in Fig. \ref{fig:ratios_maps_R2}, we computed Spearman's Rank correlation coefficients, $r_\mathrm{s}$, between the column density ratios and the H$_2$ column densities. Although the H$_2$ column density is not a perfect tracer for embeddedness, several studies have inferred an approximately linear relation between $N$(H$_2$) and $A_\mathrm{v}$ \citep{1978ApJ...224..132B, 2009MNRAS.400.2050G}. 

Figure \ref{fig:scatter} shows scatter plots of H$^{13}$CN/HC$^{15}$N and $^{14}$N/$^{15}$N column density ratios versus $N$(H$_2$) for each of the four cores. Note that the $^{14}$N/$^{15}$N ratios are derived using the double-isotope method for comparison with previous studies, which used this method. Assuming a fixed $^{12}$C/$^{13}$C = 68 \citep{2005ApJ...634.1126M}, the column density of the main HCN isotopolog was computed as $N$(HCN) = 68~$\times$~$N$(H$^{13}$CN).
 The H$_2$ column density was averaged over the same region as the HCN spectra for consistency. The Spearman correlation coefficients were computed for each core individually. 
For L694-2, the analysis shows a strong positive correlation between $N$(H$_2$) (and, consequently, $A_\mathrm{v}$ by proxy) and the $^{14}$N/$^{15}$N ratio. L429 and HMM-1 show a weaker, but still noticeable, positive correlation. No apparent correlation is present toward L1521E. The lack of correlation in L1521E can be attributed to the early evolutionary stage of the core, as evidenced by the lower $N(\text{H}_2)$ column densities toward this core with respect to the others.

\begin{figure}[ht]
\resizebox{\hsize}{!}
        {\includegraphics{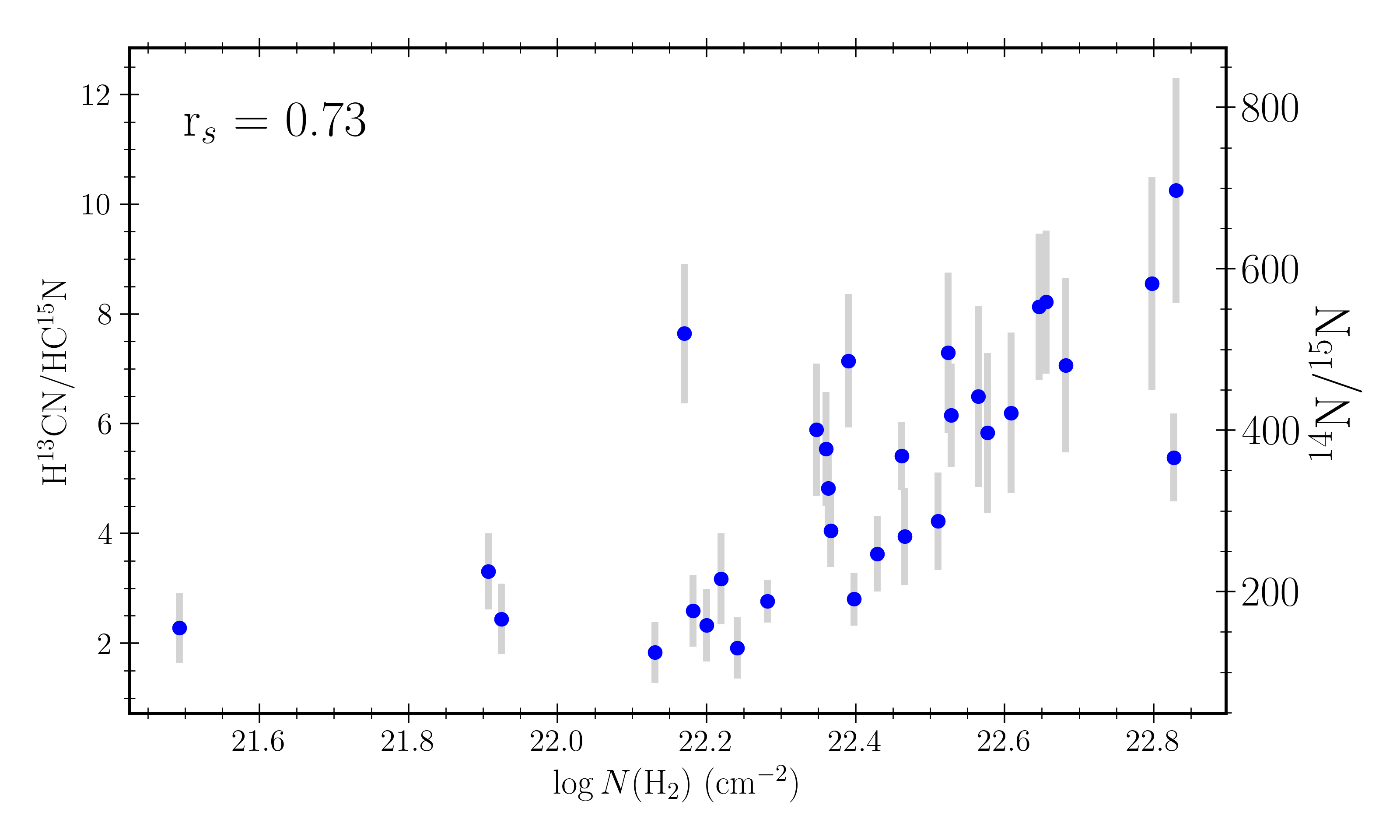}}
  \caption{The H$^{13}$CN/HC$^{15}$N column density ratio (left y-axis) and inferred $^{14}$N/$^{15}$N ratio (right y-axis) as a function of $N$(H$_2$) for all four cores. The $^{14}$N/$^{15}$N ratios were calculated by naively assuming a constant  $^{12}$C/$^{13}$C = 68 across the maps. The computed Spearman correlation coefficient is shown in the top-left corner.}
     \label{fig:scatter_all}
\end{figure}

When the points from all four cores are combined, the correlation remains high ($r_{s}\simeq0.73$, p-value = 2$\times10^{-4}$)\footnote{The p-value indicates the probability of the null hypothesis, which in this case is no correlation between the two parameters.}, indicating a general relationship between the $^{14}$N/$^{15}$N and H$^{13}$CN/HC$^{15}$N ratios and the integrated H$_2$ column density. The p-value is estimated using a permutation analysis, given the low number of samples. 
The results of the combined-source analysis are shown in Fig. \ref{fig:scatter_all}.

\subsection{Chemical modeling of nitrogen and carbon fractionation in a pre-stellar core.}
\begin{figure}[ht]
\resizebox{\hsize}{!}
        {\includegraphics{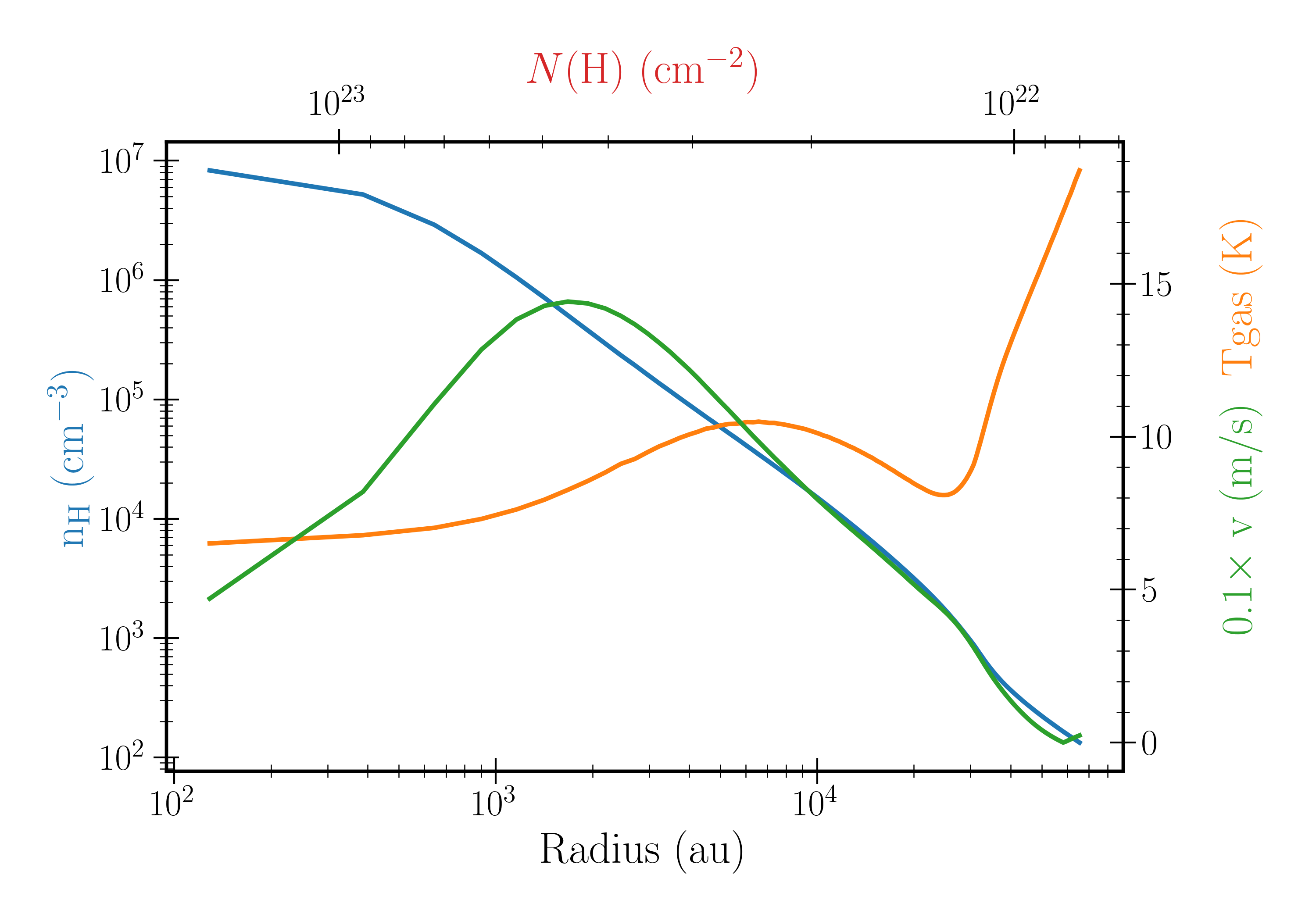}}
  \caption{Radial profiles for the number density n$_\mathrm{H}$ of total hydrogen nuclei, column density $N$(H), velocity (positive means contraction), and gas temperature for the one-dimensional L1544 model. The top axis shows the column density of total hydrogen nuclei at each radius.}
     \label{fig:core_structure}
\end{figure}

To investigate which physical and chemical processes may explain the observed spatial variations of the H$^{13}$CN/HC$^{15}$N ratios, we performed simulations using the astrochemical network presented in \citet{2023A&A...678A.120S}. The simulation was carried out on a one-dimensional model of the pre-stellar core L1544 from \citet{2015MNRAS.446.3731K}. The physical structure of the one-dimensional model is shown in Fig \ref{fig:core_structure}. 
The model was chosen because it provides a well-constrained one-dimensional representation of a pre-stellar core. It is not intended to be directly comparable to any of the cores observed in this work; rather, it serves to illustrate the temporal evolution of the $^{12}$C/$^{13}$C and $^{14}$N/$^{15}$N fractionation in an embedded, young core with a radial density profile resembling a Bonnor-Ebert sphere.

\begin{figure}[ht]
\resizebox{\hsize}{!}
        {\includegraphics{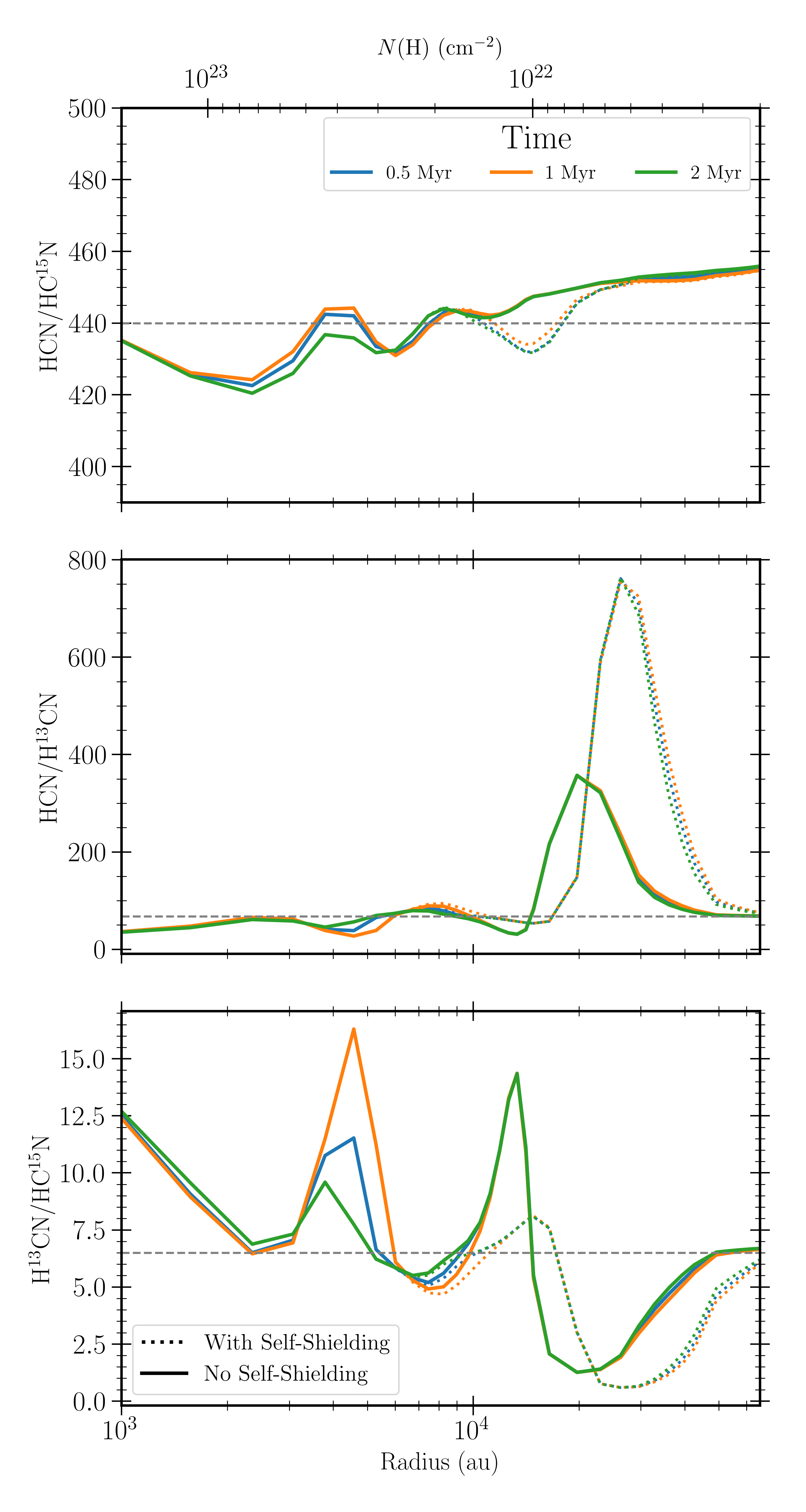}}
  \caption{Radial profiles of the HCN/HC$^{15}$N (top panel), HCN/H$^{13}$CN (middle panel), H$^{13}$CN/HC$^{15}$N (bottom panel) ratios for the one-dimensional L1544 model. The model without self-shielding is represented by the solid line, and the model with self-shielding for N$_2$ and CO is represented by the dotted lines. Horizontal dashed lines represent the ISM values for each ratio (440, 68, 6.5). The top axis shows the column density through the cloud, that is, the column density an observer would see at each impact parameter. An external extinction of $A_\mathrm{v,ext}$ = 1~mag is assumed.}
     \label{fig:radial_profiles}
\end{figure}

Figure \ref{fig:radial_profiles} shows the simulated radial profiles of the HCN/HC$^{15}$N, HCN/H$^{13}$CN, H$^{13}$CN/HC$^{15}$N ratios for a three different evolutionary times (0.5~Myr, 1~Myr, and 2~Myr). The figure includes both models with (dotted lines) and without (solid lines) the effects of isotope-selective photodissociation. 

Figure \ref{fig:radial_profiles} illustrates the impact of isotope-selective photodissociation of CO and N$_2$ on the isotopic ratios. The top panel shows that including isotope-selective photodissociation has only a limited effect on the $^{14}$N/$^{15}$N ratio; furthermore, the nitrogen fractionation ratio also exhibits little spatial variation in the one-dimensional model. In contrast, the middle panel reveals a pronounced variation of the $^{12}$C/$^{13}$C ratio. This ratio depends strongly on radial distance and also shows an additional contribution from isotope-selective photodissociation at larger radii. Moreover, the $^{12}$C/$^{13}$C ratio evolves moderately over time. 

We note that the astrochemical model has several caveats. First, the model is not dynamical and therefore cannot capture any effects arising from the dynamical formation of the core. Specifically, dynamical accretion may lead to a more pronounced impact of isotope-selective photodissociation at shorter radii, since material would be transported radially inwards as the cores evolve \citep{2018A&A...615L..16F}. 
Secondly, the isotope-selective photodissociation processes are not fully self-consistent in the current model. The self-shielding coefficients for CO/$^{13}$CO from \citet{2009A&A...503..323V} and N$_2$/$^{15}$NN from \citet{2013A&A...555A..14L} and \citet{2014A&A...562A..61H} are calculated assuming fixed $^{12}$C/$^{13}$C and $^{14}$N/$^{15}$N ratios, which especially for the former is not accurate. This may lead to an underestimation of the effects of the isotope-selective photodissociation. Furthermore, it should be noted that the radii at which the effects of isotope-selective photodissociation are efficient depend on the external extinction assumed around the core. In a three-dimensional model, with an anisotropic structure, this could lead to enhanced photodissociation deeper into the core. 
Overall, however, we do not anticipate that the stronger radial variation in $^{12}$C/$^{13}$C relative to $^{14}$N/$^{15}$N will be significantly affected by these limitations.

\section{Discussion}  \label{sec:4}
\subsection{Carbon or nitrogen fractionation?}
The maps in Figure \ref{fig:ratios_maps_R2} indicate a substantial variation in H$^{13}$CN/HC$^{15}$N column density ratios across the maps of the four cores presented here. Combined with the correlation analysis in Figs. \ref{fig:scatter} and \ref{fig:scatter_all}, this supports a general trend of higher H$^{13}$CN/HC$^{15}$N ratios toward the centers of the cores and toward the more shielded regions of the core. However, from these data alone, it is not immediately clear whether the observed gradients in the maps are the result of nitrogen or carbon fractionation.

Recent astrochemical models of carbon fractionation suggest a notable degree of carbon fractionation under cold core conditions \citep[e.g.,][]{2015A&A...576A..99R, 2020A&A...640A..51C, 2020MNRAS.498.4663L}. Indeed, using astrochemical models which treat both carbon and nitrogen fractionation, \citet{2023A&A...678A.120S} found that the carbon fractionation ratio evolves significantly over time, while nitrogen fractionation is limited ($\lesssim15\%$). This is also the case when running the same model for a one-dimensional pre-stellar core model, as presented in this work. 
In the model, the HCN/H$^{13}$CN ratio shows a stronger radial variation and a stronger impact of isotope-selective photodissociation, compared with the limited variation in the HCN/HC$^{15}$N ratio.
 This result is significant, as it directly challenges the basic assumption of the double isotope method: the $^{12}$C/$^{13}$C ratio is not stable, whereas the $^{14}$N/$^{15}$N ratio remains roughly constant, at least for HCN. 
 
 The results of these astrochemical models are supported by several observational studies. In an earlier work, using single-pointing observations, \citet{2002ApJ...575..250I} studied the H$^{13}$CN/HC$^{15}$N ratios toward the center of three young cores in Taurus, including L1521E, which is also presented here. They found different H$^{13}$CN/HC$^{15}$N ratios for each of the cores and concluded that the variation is more likely the result of varying $^{12}$C/$^{13}$C ratios, rather than $^{14}$N/$^{15}$N ratios. That conclusion was based on observations of the $^{13}$C$^{18}$O/$^{12}$C$^{18}$O ratios toward the same sources, as well as theoretical considerations, and is consistent with the results of the model presented here. More recently, a number of studies employing non-LTE radiative transfer modeling of HCN and H$^{13}$CN emission lines have derived HCN/H$^{13}$CN ratios that deviate from the ISM value of 68, typically being lower by factors of 2--3 in young cores \citep{2013A&A...560A...3D, 2018A&A...615A..52M, 2024A&A...685A.149J}.

Studies of the carbon fractionation of complex organic molecules (COMs), including CH$_3$OH and CH$_3$CN, have also shown a notable fractionation. \citet{2023A&A...669A.137H, 2025A&A...696A...1H} found a CH$_{3}^{12}$CN/CH$_{3}^{13}$CN ratio of 25--35 toward the binary protostellar system SVS13A. Similarly, \citet{2025A&A...699A.359B} found $^{12}$C/$^{13}$C in the range from 4--30 for CH$_3$OH and CH$_3$CN toward seven protostars as part of the PRODIGE program.
Comparable $^{12}$C/$^{13}$C ratios have also been found in the ALMA-PILS survey toward the protostellar multiple system IRAS16293--2422, notable for larger COMs such as glycolaldehyde (CH$_2$OHCHO), and tentatively for C$_2$H$_5$OH, CH$_3$OCHO, and CH$_3$OCH$_3$ \citep{2016A&A...595A.117J, 2018A&A...620A.170J}. 
Hence, there is growing evidence for carbon fractionation in both simpler species, such as nitriles, and in more complex molecules.

Historically, the double isotope method, namely a fixed $^{12}$C/$^{13}$C ratio, has been used extensively in the literature. Under that assumption, the observed variation would indicate higher $^{14}$N/$^{15}$N ratios in denser and more shielded parts of the core.
This is similar to the trend observed in L1544 by \citet{2022A&A...664L...2S} and is consistent with isotope-selective photodissociation playing a significant role in the nitrogen fractionation in these cores.
It is important to note that the analysis of CN and $^{13}$CN maps presented in \citet{2022A&A...664L...2S} suggests that the variation in H$^{13}$CN and HC$^{15}$N in L1544 cannot be fully explained by $^{12}$C/$^{13}$C variations alone and hence the results here do not change the overall conclusions of that paper. Additionally, \citet{2023A&A...674L...8R} reported a spatial trend in the nitrogen fractionation of NH$_3$ in L1544, consistent with the trend seen in HCN and CN. Furthermore, evidence of isotope-selective photodissociation has been observed in protoplanetary disks. \citet{2019A&A...632L..12H} reported a notable radial evolution in the HCN/HC$^{15}$N ratio for TW Hya, with a lower ratio in the inner disk, gradually increasing at larger radii. This is consistent with isotope-selective photodissociation driven by the UV radiation in the inner disk from the T Tauri star \citep{2018A&A...615A..75V}. 
 In this work, we do not have maps of CN and $^{13}$CN and hence rely on the astrochemical modeling to investigate the source of the observed variations across the maps. From Fig. \ref{fig:radial_profiles}, we see that the H$^{13}$CN/HC$^{15}$N ratio in the model is expected to increase from $\sim$5 at 30,000--60,000~au to >10 at $\sim$2,000--3,000~au. This trend is qualitatively similar to the increase seen in the observations as presented in Figs. \ref{fig:scatter} and \ref{fig:scatter_all}, although the highest values fall short of 10 in the observed data. 
 This difference may be a result of the limited spatial resolution of the observations, which sample the positions on scales of $\sim$25$^{''}$, corresponding to linear scales of $\sim$4,000--6,000~au at distances of 150~pc to 250~pc, where the sources presented here are located.
 
To summarize, the observed variations in the H$^{13}$CN/HC$^{15}$N ratio are likely driven by carbon fractionation. However, some degree of nitrogen fractionation may also be present, influencing the observed trends. To assess this, observations of a broader range of molecules are necessary to better constrain the extent of carbon and nitrogen fractionation.

\subsection{Origin of the fractionation}
By combining the data points from all four sources in Figure \ref{fig:scatter_all}, we see that the observed trend of increasing H$^{13}$CN/HC$^{15}$N ratios with $N$(H$_2$) persists when the individual cores are merged. 
This indicates that local variations in the cloud environment (e.g., variable irradiation of the sources) are not the main driver of the observed trend. The sources are located in different clouds and experience different irradiation. If variation in irradiation were the dominant factor, this variability should introduce greater scatter and tend to obscure any correlation when data points from all sources are combined. Instead, the apparent correlation in Fig. \ref{fig:scatter_all} is consistent with intrinsic effects being the main driver of the observed trend. Both carbon and nitrogen fractionation proceed through exchange reactions, which promote more efficient fractionation at lower temperatures and higher densities. Consequently, such regions will experience enhanced carbon fractionation rates, matching the observed trend with $N$(H$_2$). Indeed, the correlation in Fig. \ref{fig:scatter_all} appears to start at a cutoff of $\log_{10} (N(\mathrm{H}_2))$~$\gtrsim$~22.2. This may suggest that the pathways driving the correlation remain inefficient below this threshold, which would explain the lack of correlation observed toward L1521E, since that core does not exceed the $N$(H$_2$) threshold.
We stress that we do not exclude secondary effects driven by environmental factors, including isotope-selective photodissociation, playing a role in the fractionation in these cores. These effects may indeed be a major contributor to the scatter among points with similar $N$(H$_2$) in Fig. \ref{fig:scatter_all} and in the individual cores in Fig. \ref{fig:scatter}. To provide stronger constraints on the underlying chemistry and environmental effects, observations of additional sources and of more molecular tracers are required.

Based on the modeling presented here, the main source of the radial variation in the H$^{13}$CN/HC$^{15}$N ratio is chemical fractionation of carbon through isotopic exchange reactions. H$^{13}$CN (HC$^{15}$N) is primarily formed through electron recombination of H$^{13}$CNH$^{+}$ (HC$^{15}$NH$^{+}$) in the gas-phase. The second most important pathway is grain-surface formation through H + $^{x}$C$^{x}$N $\rightarrow$ H$^{x}$C$^{x}$N at higher densities (shorter radii). For both pathways, chemical fractionation occurs in the gas phase via various exchange reactions, thereby enriching the reactants in $^{13}$C and $^{15}$N before HCN formation.
In the one-dimensional core model, the effect of isotope-selective photodissociation is negligible at radii below $\sim$8000~au, whereas the effect is strongest at radii of $\gtrsim$~20,000~au (Fig. \ref{fig:radial_profiles}, middle panel). Compared with the observations, the modeling suggests that isotope-selective photodissociation should have only a limited impact on the correlations presented in Figs. \ref{fig:scatter} and \ref{fig:scatter_all}, since the correlation extends to column densities above $N$(H$_2$)~$ \gtrsim3\times10^{22}$~cm~$^{-2}$, where the effect is limited in the astrochemical model. However, this does not rule out isotope-selective photodissociation, since the effects may be enhanced at shorter radii in a realistic three-dimensional core with uneven irradiation, especially if dynamic accretion transports fractionated material inwards. Nonetheless, both the model and the observations show a clear trend toward higher fractionation at higher H$_2$ column densities, driven by chemical fractionation processes.
In future work, we will investigate nitrogen and carbon fractionation in a dynamical three-dimensional model of a young core embedded in a realistic star-forming environment. That work will address several caveats of the model presented here and provide a better understanding of the impact of isotope-selective photodissociation.

\begin{figure*}[ht]
\resizebox{\hsize}{!}
        {\includegraphics{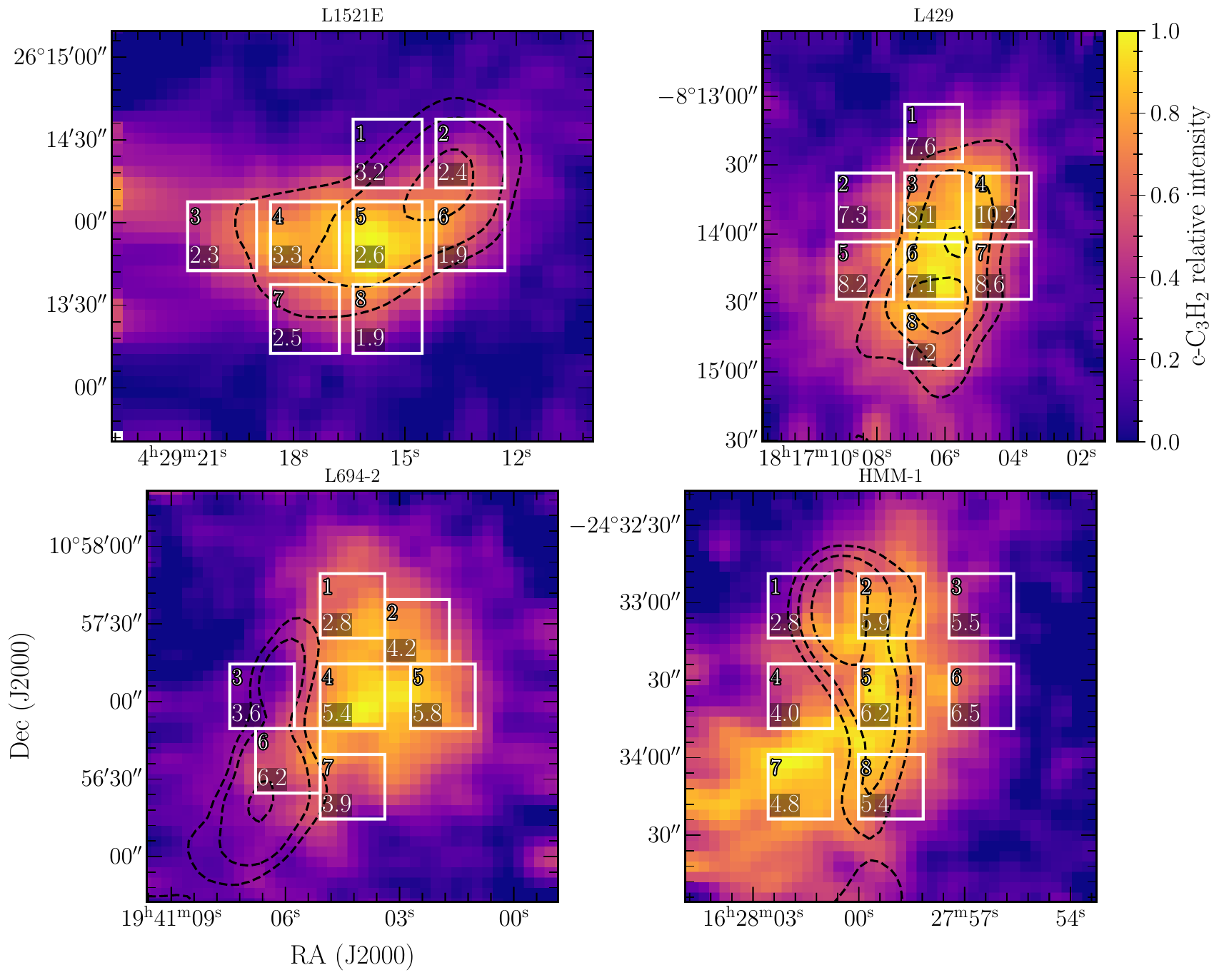}}
  \caption{Comparison of the observed H$^{13}$CN/HC$^{15}$N ratios, the $c$-C$_3$H$_2$ emission (colormap) and the CH$_3$OH emission (black contours) for the four cores. Observational $c$-C$_3$H$_2$ and the CH$_3$OH emission maps are from \citet{2020A&A...643A..60S}. The colormap is scaled to the peak $c$-C$_3$H$_2$ intensity for the specific core. Note that errors on the H$^{13}$CN/HC$^{15}$N ratios have been removed for clarity.}
     \label{fig:C3H2}
\end{figure*}

The chemical distribution of $c$--C$_3$H$_2$ and CH$_3$OH was previously studied in the cores presented in this work by \citet{2020A&A...643A..60S}. That work explored the origins of the observed molecular segregation between the two species based on the influence of external irradiation and the local structure of the core regions. In the more shielded regions, carbon is locked in CO, and CH$_3$OH is formed after CO freeze-out through grain-surface chemistry. Hence, the emission of CH$_3$OH in young cores typically traces denser and more shielded regions of the core. Meanwhile, $c$--C$_3$H$_2$ formation is enhanced through atomic carbon, which is more present in regions more exposed to the interstellar radiation field \citep{2016A&A...592L..11S, 2023A&A...675A..34J, 2026A&A...708A.279J}. 
By comparing our maps with those of $c$--C$_3$H$_2$ and CH$_3$OH presented in \citet{2020A&A...643A..60S}, we can assess whether trends in chemical segregation are consistent with those in the observed fractionation ratios. Through isotope-selective photodissociation, the HCN/HC$^{15}$N and H$^{13}$CN/HC$^{15}$N ratios should decrease in regions with lower shielding or stronger irradiation due to (sub)stellar objects \citep{2018ApJ...857..105F}. Hence, we look for an overlap in the $c$--C$_3$H$_2$ emission and lower H$^{13}$CN/HC$^{15}$N ratio. A comparison between the H$^{13}$CN/HC$^{15}$N ratios and the integrated intensity maps of $c$--C$_3$H$_2$ and CH$_3$OH is shown in Fig. \ref{fig:C3H2}. 
An apparent overlap in lower H$^{13}$CN/HC$^{15}$N ratios and stronger $c$--C$_3$H$_2$ emission is seen in HMM--1 (P4, P7). These points also coincide with the side of the core exposed to irradiation from an OB association \citep{2008hsf2.book..235P}. 
For L1521E, no clear spatial segregation was reported in \citet{2020A&A...643A..60S}. This was attributed to the core's early evolutionary stage, as indicated by the lower H$_2$ column densities toward the core. As the fractionation processes through exchange reactions are accelerated at higher column densities and lower temperatures, the low densities might also explain the apparent lack of correlation between $N$(H$_2$) and H$^{13}$CN/HC$^{15}$N in this core, seen in Fig. \ref{fig:scatter}. If isotope-selective photodissociation is active in this core, this could explain the lower H$^{13}$CN/HC$^{15}$N ratios observed across it, specifically toward the south-east of the core (P6, P8), which also shows the strongest drop in H$_2$ column densities.
For L429, $c$--C$_3$H$_2$ and CH$_3$OH emission overlap and coincide with the dust peak.
Toward L694--2, $c$--C$_3$H$_2$ emission coincides with the dust peak, while CH$_3$OH is notably offset towards the south-west direction (P6 in this work, see Fig. \ref{fig:C3H2}). This direction also shows a shallower drop in $N$(H$_2$), indicating that the south-western region is denser and more shielded from the interstellar radiation field. Given the correlation shown in Fig. \ref{fig:scatter_all}, the enhanced H$^{13}$CN/HC$^{15}$N ratio in P6 could result from the denser and more shielded conditions. This is consistent with an increased CH$_3$OH emission from this region. 
While this comparison between the $c$--C$_3$H$_2$ and CH$_3$OH maps of \citet{2020A&A...643A..60S} and this work is qualitative in nature, the similarities could suggest that photochemical effects do play a role in the fractionation morphology, even if they are not expected to be the primary mechanism. More sources and additional molecular tracers are necessary to shed more light on this.

\section{Conclusions}  \label{sec:5}
In this work, we have studied the spatial variation in the nitrogen and carbon fractionation around four young, embedded cores. The observations reveal a correlation: regions with higher H$_2$ column densities exhibit larger H$^{13}$CN/HC$^{15}$N column density ratios. This corresponds to an increase in the $^{14}$N/$^{15}$N ratio when the double isotope method is applied. The correlation also persists when data across all four cores are combined.
A priori, two mechanisms can generate the observed trend: i) enhanced self-shielding of N$_2$ or CO leading to isotope-selective photodissociation effects which influence the fractionation processes, or ii) increased fractionation through isotopic exchange reactions, which are more efficient at higher densities and lower temperatures.

To distinguish between these scenarios, we simulated carbon and nitrogen fractionation using a one-dimensional, time-dependent astrochemical model of a pre-stellar core. The model predicts limited nitrogen fractionation at all times and radii, whereas carbon fractionation varies substantially as a function of both radius and time. This is in direct contrast to the fundamental assumption of the double isotope method, namely a fixed $^{12}$C/$^{13}$C ratio. Isotope-selective photodissociation remains modest in the astrochemical model and is limited to the outer parts of the cores.
Instead, chemical fractionation through isotopic exchange reactions is the primary driver of fractionation.
We note, however, that a full three-dimensional treatment that includes dynamical accretion, anisotropic radiation fields, and realistic envelope structure could amplify the effects of isotope-selective photodissociation, especially in strongly irradiated environments.

The observed correlation between the $N$(H$^{13}$CN)/$N$(HC$^{15}$N) ratios and $N$(H$_2$) suggests that the fractionation of carbon in these young cores follows a fairly continuous evolution above a threshold of $\log_{10} (N(\mathrm{H}_2))$~$\gtrsim$~22.2, with increasing fractionation at higher densities and lower temperatures, in line with the expectation for chemical fractionation.

Our results, along with other recent studies, indicate that the fundamental assumption underlying the double isotope method is incorrect. Hence, we caution future work not to rely on the double isotope method when possible. This caution not only applies to young cores, but should also be considered at later stages of the star and planet formation process, where the assumption of a spatially and temporally invariant $^{12}$C/$^{13}$C ratio may lead to systematic errors.
 
Future observations, including additional molecules (e.g., HNC, CN, and complex organic molecules) toward more sources, would provide stronger constraints on the observed correlation. In addition, more comprehensive modeling, including dynamical accretion and a realistic three-dimensional structure, would provide a clearer understanding of how much this mechanism influences the fractionation process in these cores.

\begin{acknowledgements}
The authors wish to thank the anonymous referee for their comments which improved the manuscript.
S.S.J. and S.S. wish to thank the Max Planck Society for the Max Planck Research Group funding. O.S. and P.C. gratefully acknowledge the funding by the Max
Planck Society. The project that gave rise to these results received the support of a fellowship from the ``la Caixa'' Foundation (ID 100010434). The fellowship code is LCF/BQ/PR25/12110012. L.C. acknowledges funding from grant PID2022-136814NB-I00 funded by the Spanish Ministry of Science, Innovation and Universities/State Agency of Research MICIU/AEI/ 10.13039/501100011033 and by ``ERDF/EU''.
This work is based on observations carried out under project numbers 117-22, 121-24, and 023-25 with the IRAM NOEMA Interferometer [30m telescope]. IRAM is supported by INSU/CNRS (France), MPG (Germany) and IGN (Spain). We wish to thank the telescope staff for their superb assistance during the observations.
 This paper makes use of {\sc matplotlib} \citep{Hunter:2007} and {\sc scipy} \citep{2020SciPy-NMeth}.
\end{acknowledgements}

%
%

\bibliographystyle{aa}
\bibliography{n.bib}

\begin{appendix}
\section{Extracted spectra}\label{app:figs}
\begin{figure*}[ht]
\resizebox{\hsize}{!}
        {\includegraphics{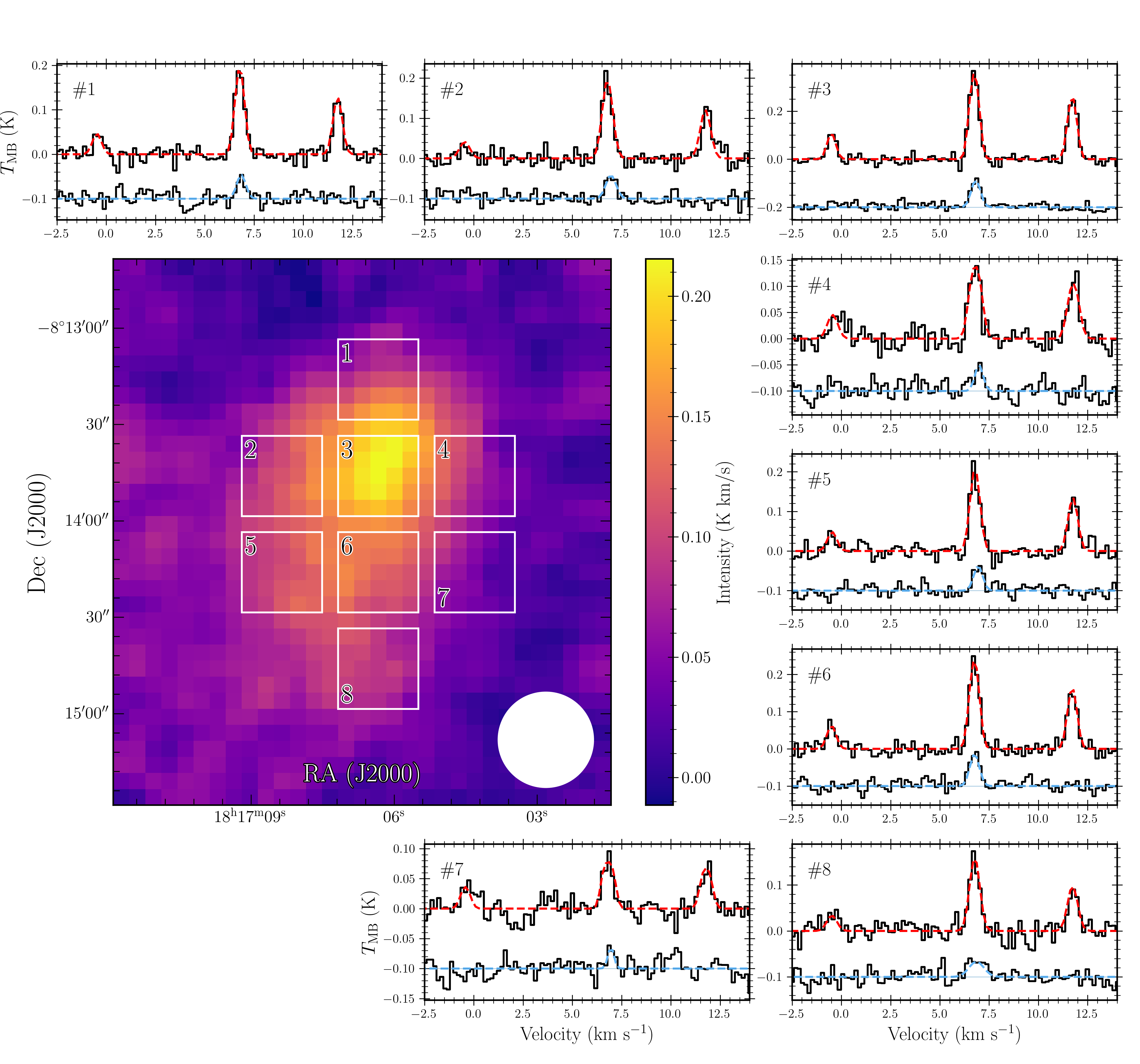}}
  \caption{Moment-zero map toward L429 for the central hyperfine component of H$^{13}$CN (1--0). White rectangles indicate the regions over which the spectra are averaged. The H$^{13}$CN and HC$^{15}$N spectra for each of the 8 positions are plotted in black, the latter offset for clarity. The hyperfine fit for H$^{13}$CN is shown in red, and the Gaussian fit for HC$^{15}$N is shown in blue. The beam size is shown in the lower right corner.}
     \label{fig:L429}
\end{figure*}

\begin{figure*}[ht]
\resizebox{\hsize}{!}
        {\includegraphics{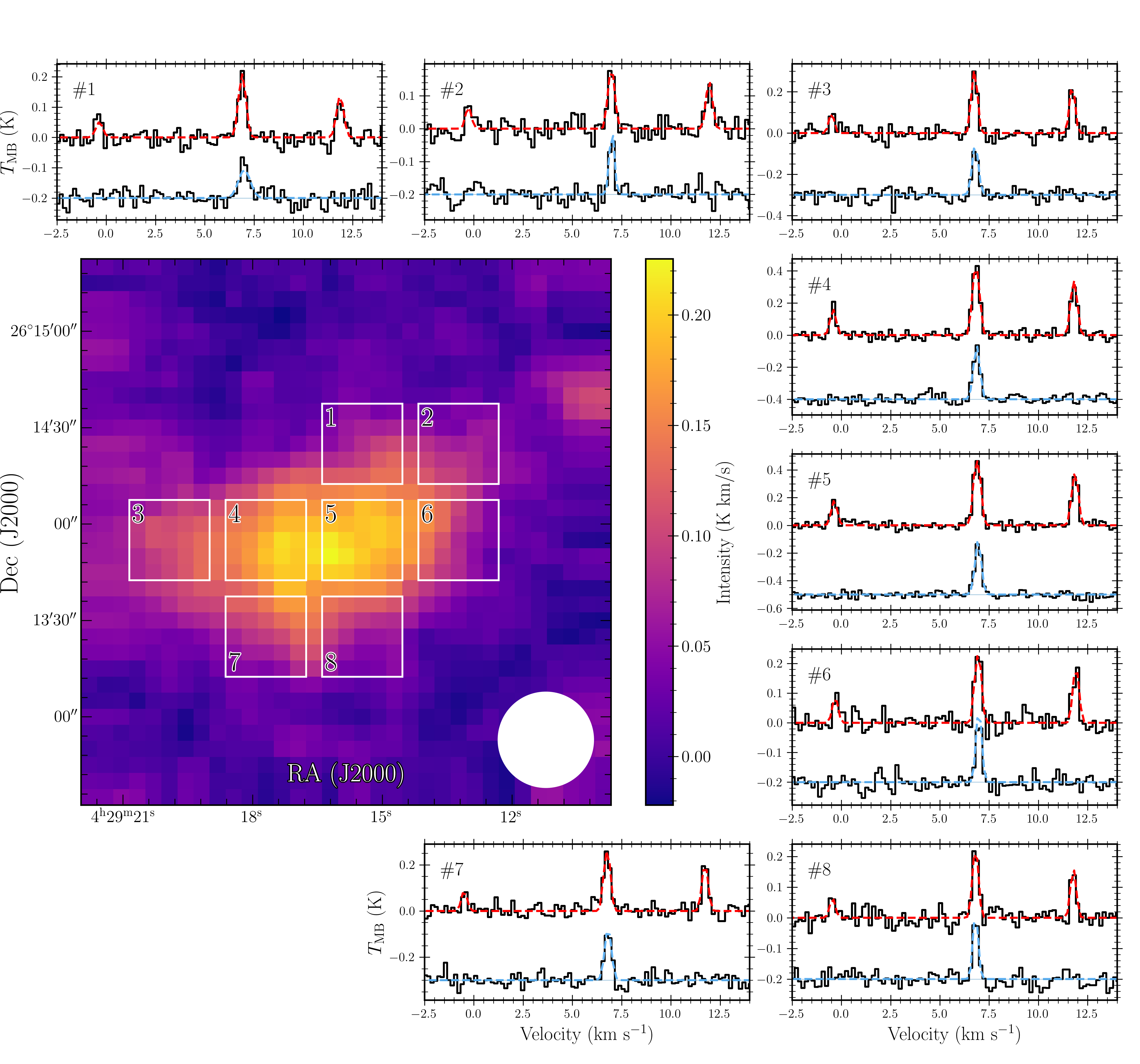}}
  \caption{Moment-zero map toward L1521E for the central hyperfine component of H$^{13}$CN (1--0). White rectangles indicate the regions over which the spectra are averaged. The H$^{13}$CN and HC$^{15}$N spectra for each of the 8 positions are plotted in black, the latter offset for clarity. The hyperfine fit for H$^{13}$CN is shown in red, and the Gaussian fit for HC$^{15}$N is shown in blue. The beam size is shown in the lower right corner.}
     \label{fig:L1521E}
\end{figure*}

\begin{figure*}[ht]
\resizebox{\hsize}{!}
        {\includegraphics{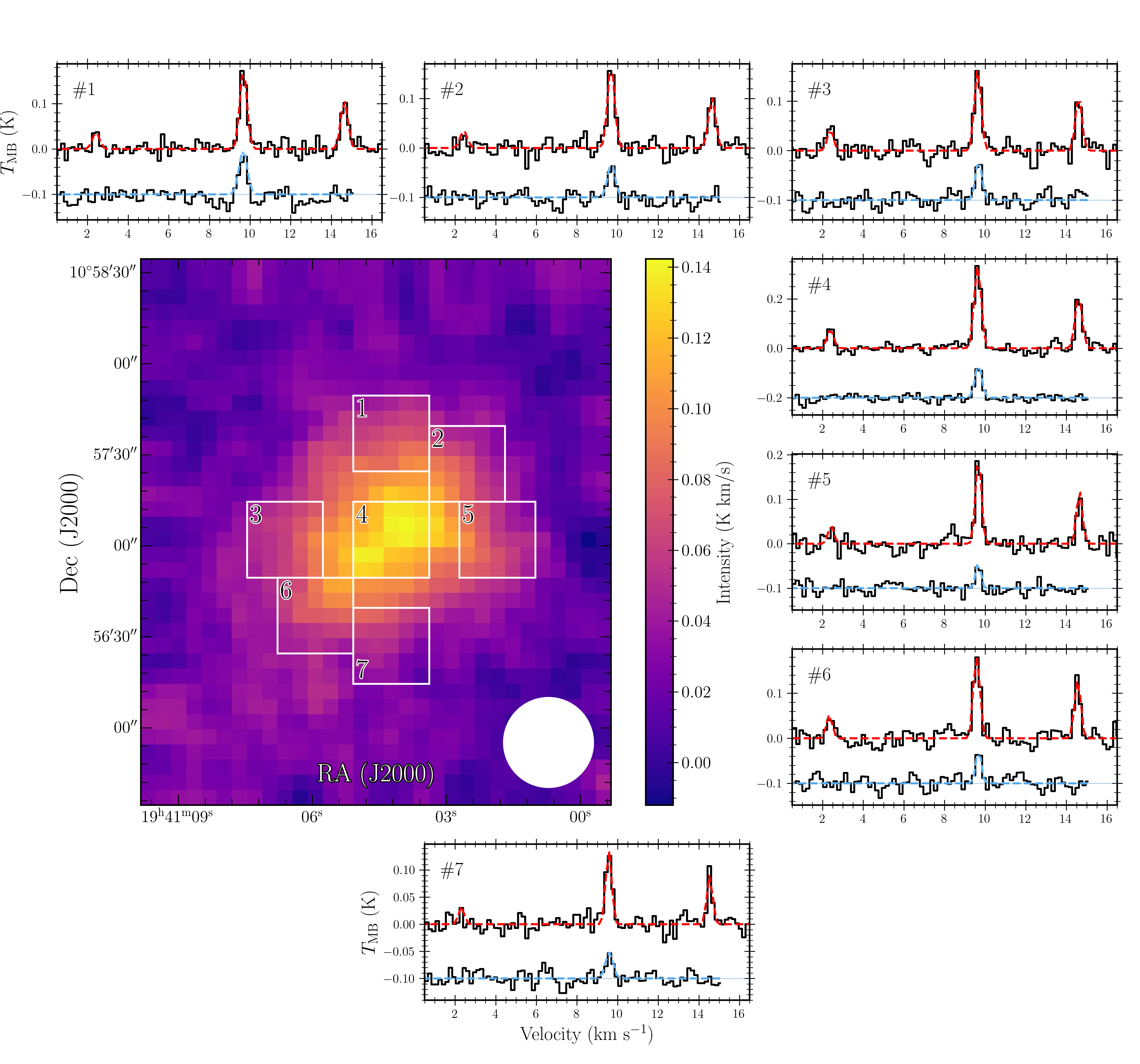}}
  \caption{Moment-zero map toward L694--2 for the central hyperfine component of H$^{13}$CN (1--0). White rectangles indicate the regions over which the spectra are averaged. The H$^{13}$CN and HC$^{15}$N spectra for each of the 8 positions are plotted in black, the latter offset for clarity. The hyperfine fit for H$^{13}$CN is shown in red, and the Gaussian fit for HC$^{15}$N is shown in blue. The beam size is shown in the lower right corner.}
     \label{fig:L694-2}
\end{figure*}

Figures \ref{fig:L429}-\ref{fig:L694-2} show the extracted spectra and fitted profiles for L429, L1521E, and L694--2, similar to Fig. \ref{fig:HMM-1}.

\FloatBarrier
\section{Overview of derived properties.}\label{app:tables}
Tables \ref{table:L694--2}-\ref{table:HMM1} list the derived column densities and optical depths $\tau$ for each position toward the four cores. For H$^{13}$CN, the listed optical depth corresponds to the component used to derive the column density. One source, L1521E, shows notably higher optical depth for both H$^{13}$CN and HC$^{15}$N. 

\begin{table*}
\centering\caption{Derived properties toward L694--2.}             
\label{table:L694--2}
\centering          
\begin{tabular}{lcccc}
\hline
\hline
Position & $N(\mathrm{H^{13}CN})$ (cm$^{-2}$) & $N(\mathrm{HC^{15}N})$ (cm$^{-2}$) & $\tau_{\mathrm{H^{13}CN}}$ & $\tau_{\mathrm{HC^{15}N}}$ \\ 
\hline
P1 & (7.0 $\pm$ 0.6) $\times 10^{11}$ & (2.5 $\pm$ 0.4) $\times 10^{11}$ & $0.09 \pm 0.01$ & $0.08 \pm 0.01$ \\ 
P2 & (6.7 $\pm$ 0.6) $\times 10^{11}$ & (1.6 $\pm$ 0.3) $\times 10^{11}$ & $0.09 \pm 0.01$ & $0.06 \pm 0.01$ \\ 
P3 & (5.8 $\pm$ 0.5) $\times 10^{11}$ & (1.6 $\pm$ 0.3) $\times 10^{11}$ & $0.10 \pm 0.01$ & $0.07 \pm 0.01$ \\ 
P4 & (1.6 $\pm$ 0.1) $\times 10^{12}$ & (2.9 $\pm$ 0.4) $\times 10^{11}$ & $0.08 \pm 0.01$ & $0.12 \pm 0.01$ \\ 
P5 & (5.3 $\pm$ 0.5) $\times 10^{11}$ & (9.0 $\pm$ 2.0) $\times 10^{10}$ & $0.09 \pm 0.01$ & $0.05 \pm 0.01$ \\ 
P6 & (7.6 $\pm$ 0.7) $\times 10^{11}$ & (1.2 $\pm$ 0.3) $\times 10^{11}$ & $0.12 \pm 0.01$ & $0.06 \pm 0.01$ \\ 
P7 & (4.3 $\pm$ 0.4) $\times 10^{11}$ & (1.1 $\pm$ 0.2) $\times 10^{11}$ & $0.09 \pm 0.01$ & $0.04 \pm 0.01$ \\ 
\hline
\end{tabular}
\end{table*}

\begin{table*}
\centering\caption{Derived properties toward L1521E.}             
\label{table:L1521E}
\centering          
\begin{tabular}{lcccc}
\hline
Position & $N(\mathrm{H^{13}CN})$ (cm$^{-2}$) & $N(\mathrm{HC^{15}N})$ (cm$^{-2}$) & $\tau_{\mathrm{H^{13}CN}}$ & $\tau_{\mathrm{HC^{15}N}}$ \\ 
\hline
P1 & (4.1 $\pm$ 0.6) $\times 10^{12}$ & (1.3 $\pm$ 0.3) $\times 10^{12}$ & $0.8 \pm 0.1$ & $0.3 \pm 0.2$ \\ 
P2 & (3.4 $\pm$ 0.4) $\times 10^{12}$ & (1.5 $\pm$ 0.4) $\times 10^{12}$ & $0.4 \pm 0.1$ & $0.8 \pm 0.4$ \\ 
P3 & (5.6 $\pm$ 0.9) $\times 10^{12}$ & (2.5 $\pm$ 0.5) $\times 10^{12}$ & $0.92 \pm 0.1$ & $1.0 \pm 0.3$ \\ 
P4 & (1.10 $\pm$ 0.04) $\times 10^{13}$ & (3.3 $\pm$ 0.7) $\times 10^{12}$ & $0.61 \pm 0.02$ & $1.3 \pm 0.4$ \\ 
P5 & (1.02 $\pm$ 0.03) $\times 10^{13}$ & (3.9 $\pm$ 0.6) $\times 10^{12}$ & $0.48 \pm 0.01$ & $1.6 \pm 0.4$ \\ 
P6 & (4.5 $\pm$ 0.6) $\times 10^{12}$ & (2.3 $\pm$ 0.5) $\times 10^{12}$ & $0.6 \pm 0.1$ & $1.1 \pm 0.4$ \\ 
P7 & (5.6 $\pm$ 0.8) $\times 10^{12}$ & (2.3 $\pm$ 0.4) $\times 10^{12}$ & $0.8 \pm 0.1$ & $0.9 \pm 0.2$ \\ 
P8 & (3.2 $\pm$ 0.4) $\times 10^{12}$ & (1.7 $\pm$ 0.4) $\times 10^{12}$ & $0.5 \pm 0.1$ & $0.9 \pm 0.4$ \\ 
\hline
\end{tabular}
\end{table*}

\begin{table*}
\centering\caption{Derived properties toward L429.}             
\label{table:L429}
\centering          
\begin{tabular}{lcccc}
\hline
Position & $N(\mathrm{H^{13}CN})$ (cm$^{-2}$) & $N(\mathrm{HC^{15}N})$ (cm$^{-2}$) & $\tau_{\mathrm{H^{13}CN}}$ & $\tau_{\mathrm{HC^{15}N}}$ \\ 
\hline
P1 & (3.5 $\pm$ 0.6) $\times 10^{12}$ & (4.6 $\pm$ 0.5) $\times 10^{11}$ & $0.34 \pm 0.04$ & $0.14 \pm 0.05$ \\ 
P2 & (4.0 $\pm$ 0.8) $\times 10^{12}$ & (5.5 $\pm$ 0.6) $\times 10^{11}$ & $0.3 \pm 0.1$ & $0.14 \pm 0.02$ \\ 
P3 & (8.5 $\pm$ 1.3) $\times 10^{12}$ & (1.1 $\pm$ 0.1) $\times 10^{12}$ & $0.28 \pm 0.03$ & $0.28 \pm 0.04$ \\ 
P4 & (3.9 $\pm$ 0.6) $\times 10^{12}$ & (3.8 $\pm$ 0.6) $\times 10^{11}$ & $0.29 \pm 0.03$ & $0.14 \pm 0.04$ \\ 
P5 & (4.3 $\pm$ 0.7) $\times 10^{12}$ & (5.2 $\pm$ 0.6) $\times 10^{11}$ & $0.6 \pm 0.1$ & $0.15 \pm 0.03$ \\ 
P6 & (5.5 $\pm$ 1.2) $\times 10^{12}$ & (7.8 $\pm$ 0.9) $\times 10^{11}$ & $0.3 \pm 0.1$ & $0.21 \pm 0.04$ \\ 
P7 & (1.7 $\pm$ 0.3) $\times 10^{12}$ & (2.0 $\pm$ 0.3) $\times 10^{11}$ & $0.21 \pm 0.02$ & $0.10 \pm 0.20$ \\ 
P8 & (3.2 $\pm$ 0.5) $\times 10^{12}$ & (4.5 $\pm$ 0.5) $\times 10^{11}$ & $0.44 \pm 0.05$ & $0.08 \pm 0.02$ \\ 
\hline
\end{tabular}
\end{table*}

\begin{table*}
\centering\caption{Derived properties toward HMM--1.}             
\label{table:HMM1}
\centering          
\begin{tabular}{lcccc}
\hline
Position & $N(\mathrm{H^{13}CN})$ (cm$^{-2}$) & $N(\mathrm{HC^{15}N})$ (cm$^{-2}$) & $\tau_{\mathrm{H^{13}CN}}$ & $\tau_{\mathrm{HC^{15}N}}$ \\ 
\hline
P1 & (1.1 $\pm$ 0.2) $\times 10^{12}$ & (3.9 $\pm$ 0.5) $\times 10^{11}$ & $0.29 \pm 0.03$ & $0.24 \pm 0.09$ \\ 
P2 & (4.3 $\pm$ 0.9) $\times 10^{12}$ & (7.3 $\pm$ 0.8) $\times 10^{11}$ & $0.3 \pm 0.2$ & $0.31 \pm 0.05$ \\ 
P3 & (1.5 $\pm$ 0.3) $\times 10^{12}$ & (2.7 $\pm$ 0.3) $\times 10^{11}$ & $0.25 \pm 0.03$ & $0.14 \pm 0.05$ \\ 
P4 & (1.4 $\pm$ 0.3) $\times 10^{12}$ & (3.4 $\pm$ 0.4) $\times 10^{11}$ & $0.27 \pm 0.03$ & $0.15 \pm 0.03$ \\ 
P5 & (5.5 $\pm$ 0.9) $\times 10^{12}$ & (9.0 $\pm$ 1.0) $\times 10^{11}$ & $0.30 \pm 0.03$ & $0.38 \pm 0.06$ \\ 
P6 & (2.1 $\pm$ 0.5) $\times 10^{12}$ & (3.3 $\pm$ 0.4) $\times 10^{11}$ & $0.22 \pm 0.05$ & $0.16 \pm 0.07$ \\ 
P7 & (1.8 $\pm$ 0.3) $\times 10^{12}$ & (3.7 $\pm$ 0.4) $\times 10^{11}$ & $0.26 \pm 0.03$ & $0.11 \pm 0.05$ \\ 
P8 & (3.6 $\pm$ 0.5) $\times 10^{12}$ & (6.5 $\pm$ 0.7) $\times 10^{11}$ & $0.46 \pm 0.05$ & $0.21 \pm 0.03$ \\ 
\hline
\end{tabular}
\end{table*}
\FloatBarrier
\section{$^{14}$N/$^{15}$N maps derived from double isotope method.}\label{app:double}
\begin{figure*}[ht]
\resizebox{\hsize}{!}
        {\includegraphics{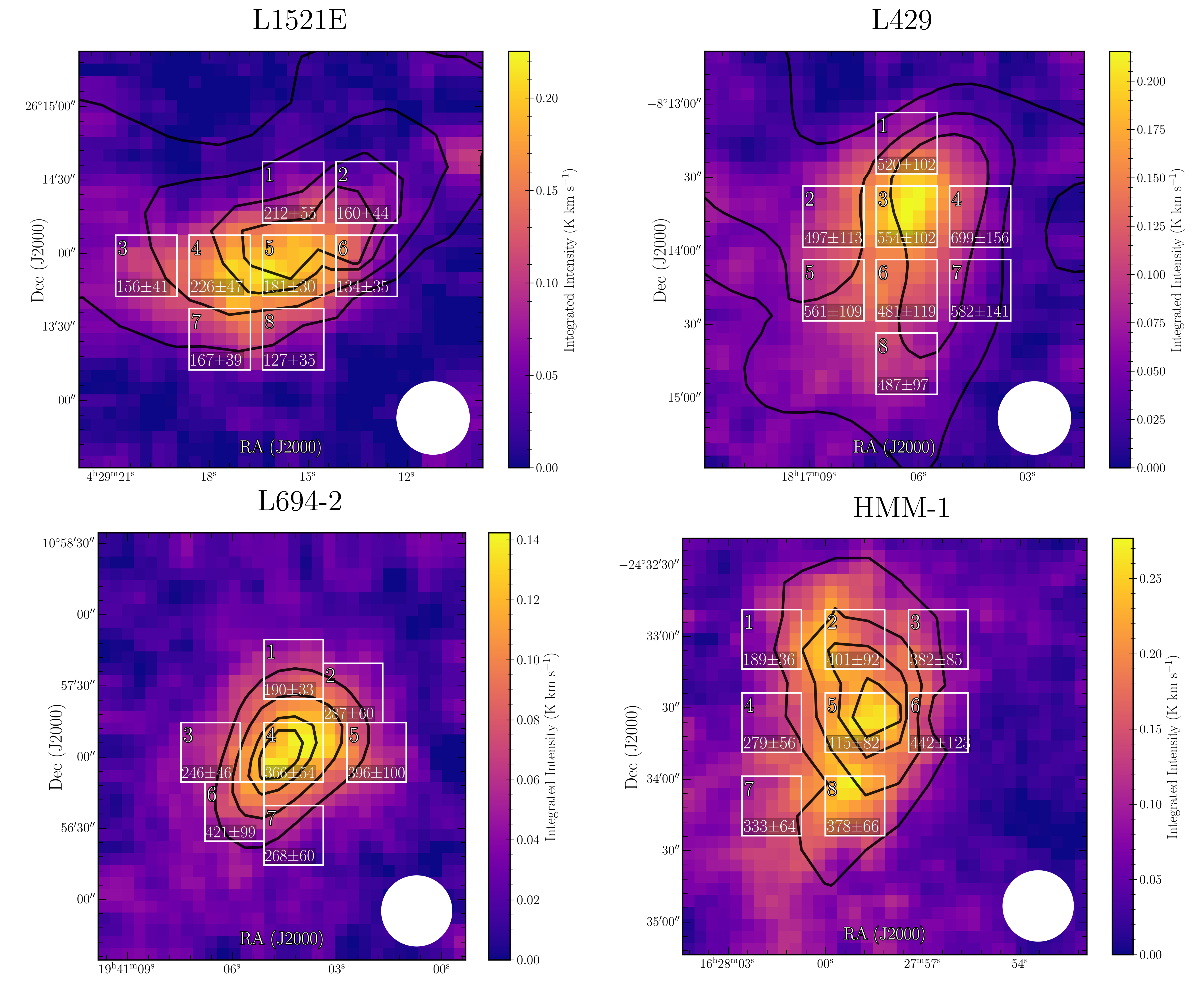}}
  \caption{Derived $^{14}$N/$^{15}$N ratios, assuming a fixed $^{12}$C/$^{13}$C ratio of 68. Black contours indicate the 10$^{22}$~cm$^{-2}$, $2.5\times$10$^{22}$~cm$^{-2}$, and $5\times$10$^{22}$~cm$^{-2}$ levels in the H$_2$ column density maps derived from \emph{Herschel}/SPIRE observations. The colormap shows the integrated emission of the central hyperfine transition for H$^{13}$CN (1--0). For L694--2, positions P2 and P6 are shifted closer to the core center to achieve sufficient S/N.}
     \label{fig:ratios_maps}
\end{figure*}

Figure \ref{fig:ratios_maps} presents maps of the $^{14}$N/$^{15}$N ratios for the four cores under the assumption of the double isotope method. 

\end{appendix}

\end{document}